# Rotational Flow Dominates Abrupt Seasonal Change in Zonally Asymmetric Tropical Meridional Circulation

Wuqiushi Yao[1,3,#], Jianhua Lu[2,#,*], Yimin Liu[1,3,*]

[1] Key Laboratory of Earth System Numerical Modeling and Application, Institute of Atmospheric Physics, Chinese Academy of Sciences, Beijing, China
[2] School of Atmospheric Sciences, Sun Yat-sen University (SYSU), Zhuhai, China
[3] College of Earth and Planetary Sciences, University of Chinese Academy of Sciences, Beijing, China

# shared first authorship;

* Corresponding to: Jianhua Lu, lvjianhua@mail.sysu.edu.cn;
              Yimin Liu, lym@lasg.iap.ac.cn

**Abstract**

The seasonality of the tropical meridional circulation evolves differently across different regions, governs the onset and retreat of monsoons and migration of tropical precipitation, thereby influencing agricultural productivity and disaster preparedness in the tropics and subtropics. By defining a pseudo meridional overturning streamfunction ($\Psi_{pseudo}$) and defining a new vector-type, dual-component index (ASCI), we diagnose zonally asymmetric abrupt seasonal change (ASC) of tropical meridional circulation. $\Psi_{pseudo}$ converges to traditional, meridional overturning streamfunction ($\Psi_m$) after being averaged over a zonal circle around any latitude. By applying the Helmholtz decomposition to horizontal velocity fields so as to decompose $\Psi_{pseudo}$ into rotational and divergent components, we quantitatively compare the contributions of horizontally rotational and divergent flows to the abrupt seasonal change. We find that the zonal sectors associated with strong deep convection exhibit the most pronounced ASC of tropical meridional circulation, and all of subregions exhibiting ASC contain landmass with low heat inertia. Particularly, in contrast to the case of zonally symmetric Hadley cell, rotational flow, rather than the thermal-direct divergent flow, dominates the zonally asymmetric ASC in the tropics, although the divergent flow also contributes to the ASC over the zonal sectors associated with deep convection. We suggest that the interplay between tropical Rossby-type eddies with extratropical eddies and tropical circulation is essential to the zonally asymmetric ASC of tropical Hadley circulation.

Keywords: Hadley circulation; Zonally asymmetric; abrupt seasonal change; Helmholtz decomposition; rotational flow

# 1 Introduction

The atmosphere circulation doesn't simply evolve gradually with season (Schneider and Bordoni, 2008). Yeh et al. (1959) discovered with limited data, that at five longitudinal spans ranging from 80 °W to 165°E of the Northern Hemispheric subtropics, the seasonal transition of zonal wind from easterly to westerly (or vice versa) is not gradual but abrupt, occurring within 5-10 days. They further suggested the abrupt seasonal change (ASC hereafter) might be a global phenomenon and proposed an idealized experiment to explore its mechanism (Lu and Schneider, 2017).

A primary emphasis on ASC has been directed toward the zonally symmetric Hadley Circulation (Bordoni and Schneider, 2008), due to its relevance to ITCZ precipitation (Schneider et al., 2014), monsoon outburst (Wang et al., 2004; Schneider and Bordoni, 2008) and subtropical drought. Dima and Wallace (2003) found that the observed seasonal change of zonally averaged Hadley Cell's strength showed no sign of abrupt change. Hu (2007) found, however, that the seasonal migration of ITCZ precipitation maxima underwent abrupt meridional jump, with abrupt change in the width of the Hadley Cell, and Xian and Miller (2008) showed similar results with ITCZ precipitation. Aqua-planet GCM experiments then substantiated this concept by showing the abrupt change of precipitation maxima, lower-level upward mass flux and the meridional position of maximum and minimum streamfunction (Schneider and Bordoni, 2008; Bordoni and Schneider, 2008). On the other hand, it is found zonally asymmetric subtropical forcing may also affect the seasonal transition of zonal-mean tropical Hadley circulation (Shaw, 2014).

Indeed, the finding of Dima and Wallace (2003) does not necessarily contradict Yeh et al.'s suggestion that ASC is a global phenomenon, because the timing of abrupt change differs longitudinally, and hence the abruptness may disappear during zonally averaging the observed data (Hu et al., 2007; Adam et al., 2016). This highlights the need for a method that may quantitatively and effectively reveal the ASC at regional scale.

Theoretical analyses based on zonal-mean angular momentum conservation and balance in zonally symmetric aquaplanet simulations have also been used to interpret the dynamics of the ACS of tropical meridional circulation that may be zonally asymmetric such as the Indian monsoon. Bordoni and Schneider (2008) point out the ASC of monsoon depends on the feedback between large-scale extratropical eddies and tropical circulation, and the rapidness of the transition, between an regime with eddy-driven momentum fluxes (equinox regime) and a more thermal-driving regime nearly conserving angular momentum (solstice regime), relies more on the small inertia of landmass than on the land-sea contrast.

Because the meridional flow (*v*) in a zonally symmetric Hadley cell contains only the divergent part ($\frac{\partial v}{\partial x} = 0, \frac{\partial v}{\partial y} \neq 0$) while the rotational flow is only contained in the zonal flow ($\frac{\partial u}{\partial x} = 0, \zeta = -\frac{\partial u}{\partial y} \neq 0$), it is natural to define in a same way the local or regional meridional circulation, or modified Hadley circulation as Lorenz (1969) call it. This also fits the expectation that a meridional overturning streamfunction should be closed at the upper and lower boundaries by fulfilling $\frac{\partial v}{\partial y} + \frac{\partial \omega}{\partial p} = 0$. Indeed, this has long been a common practice as shown in Schwendike et al. (2014), Nguyen et al.

(2018), Geen et al. (2020), and Galanti et al. (2022).

However, the rotational meridional flow ($v^{rot}$) may consist of a large part of meridional circulation at local and regional scales. Dynamically, the local and regional $v^{rot}$ are mainly associated with stationary, quasi-stationary, and transient Rossby-type eddies in both the extratropics and the tropics. Indeed we shall show in the follow sections that, by excluding the rotational flow, the ASC of zonally asymmetric divergent Hadley cell may be misleadingly too weak. Therefore, to correctly understand the rotational contribution to the meridional circulation, a modification on the definition of meridional streamfunction is necessary, and a new method is also needed.

In this paper, we attempt to develop such a method. (1) First, we define a pseudo meridional overturning streamfunction ($\Psi_{pseudo}$) at any zonal sector, which converges to traditional meridional overturning streamfunction ($\Psi_m$) after zonally averaging around any latitude. (2) Then, by dividing the zonal circle into zonal regions based on zonal-sector-averaged OLR, we define a vector-type two-component "Abrupt Seasonal Change Index" (ASCI) to describe ASC of tropical meridional circulation. (3) To quantitatively compare the contributions of rotational and divergent flows, we apply Helmholtz decomposition to horizontal velocity fields, so as to decompose $\Psi_{pseudo}$ into the sum of its rotational and divergent parts. The layout of the paper is as follows. In Section 2 we describe the datasets used in this paper. Section 3 is about the definition of $\Psi_{pseudo}$ and its Helmholtz decomposition, the ASCI which is calculated based on regional $\Psi_{pseudo}$ over the Tropics. Subsequently in Section 4 we present and analyze the ASC of meridional circulation over different tropical regions. Section 5 is about the rotational and divergent contributions to ASC. Section 6 is conclusions and discussion.

**2 Data**

We use ERA5 (horizontal resolution 1°) and JRA-55 (horizontal resolution 1.25°) reanalysis dataset at 00, 06, 12, and 18 UTC on pressure levels from 1979 to 2018 to study ASC in the tropical meridional circulation. Variables include horizontal ($u$, $v$) and vertical ($\omega$) velocities at pressure levels from 1000hPa to 10hPa. We also average the 6-hourly data into their pentad means.

We use OLR with horizontal resolution (1.0°) in Clouds and the Earth's Radiant Energy System (CERES) Energy Balanced And Filled (EBAF, Loeb et al., 2001) from NASA during 2000-2024 to represent large-scale tropical convection and to serve as the basis of region division in this study.

**3 Methods**

**3.1 Division of zonal subregions**

Similar to Hoskins et al. (2020) and Hoskins and Yang (2021), we divide the tropical areas (Table S1) according to the climatological OLR (Figure 1a). Climatologically, strong deep convection in the tropics is primarily concentrated in specific regions, namely zonal sectors containing central Africa, the South China Sea, the Maritime Continent, and South America (designated as Regions 1, 3, 4, and 7, respectively, in Figure 1a). Conversely, over vast open ocean regions such as the central-to-eastern Pacific and the Atlantic, shallow convection prevails over the zonally elongated ITCZ. Exceptions to these patterns are observed over the Indian Ocean (Region 2) and mid-tropical Pacific (Region 5), which represent the transition region in-between the deep and shallow convection regions. As such, we categorize the entire zonal circle into sub-regions of deep convection, shallow convection, and transitional regions (also see Table S1 in which the naming of sub-regions is simply for the convenience of readers in recognizing the geographic location of the sub-regions).

**3.2 Regional pseudo meridional streamfunction ($\Psi_{\text{pseudo}}$) and its decomposition into rotational and divergent parts**

The pseudo meridional streamfunction ($\Psi_{pseudo}$) at a given pressure level ($p$) and a given latitude ($\varphi$) is defined as the mass-weighted vertical integration of meridional wind ($v$) from a top pressure level to $p$

$$\Psi_{pseudo}(\varphi, p) = \int_{p_t}^{p} v \cos\varphi \, dp$$

in which $p_t$ is the pressure of the top of the atmosphere (TOA), ideally $p_t = 0$ hPa, but practically we choose $p_t = 100$ hPa with little changes in the results. **We call it "pseudo" because its value at the surface ($p_s$) is not necessarily zero at a given longitude or longitudinal sector due to the net mass flux across the lateral boundaries of the zonal sector**, but note its physical interpretation as downwardly integrated meridional mass transport is well defined. Meanwhile it is easily found that its zonal average over latitude $\varphi$ converges to the traditional meridional streamfunction $\Psi$.

From continuity equation in a spherical coordinate:

$$\frac{1}{a\cos\varphi}\left(\frac{\partial u}{\partial \lambda} + \frac{\partial v \cos\varphi}{\partial \varphi}\right) + \frac{\partial \omega}{\partial p} = 0 \qquad (1)$$

and averaging it from longitude $\lambda_1$ to $\lambda_2$,

$$\frac{1}{a\cos\varphi}\frac{\partial[v]^*\cos\varphi}{\partial\varphi}+\frac{\partial[\omega]^*}{\partial p}=RES \qquad (2)$$

In which $[X]^*=\frac{\int_{\lambda_1}^{\lambda_2}Xd\lambda}{\int_{\lambda_1}^{\lambda_2}d\lambda}$ and $RES=-\frac{1}{a\cos\varphi}\frac{u_2-u_1}{\lambda_2-\lambda_1}$.

We have a pseudo-streamfunction zonally averaged between $\lambda_1$ and $\lambda_2$:

$$[\Psi_{pseudo}]^*(\varphi,p)=\int_{p_t}^{p}[v]^*\cos\varphi dp \qquad (3)$$

in which $p_t$=100hPa is used as the top pressure level.

We further decompose the horizontal velocity field into rotational and divergent parts (Helmholtz's theorem)

$$(u,v)=(u^{rot},v^{rot})+(u^{div},v^{div}) \qquad (4)$$

Accordingly, $\Psi_{pseudo}$ can be decomposed into the sum of its rotational and divergent parts

$$\Psi_{pseudo}(\varphi,p)=\int_{p_t}^{p}[v^{rot}]^*\cos\varphi dp+\int_{p_t}^{p}[v^{div}]^*\cos\varphi dp=$$

$$\Psi_{pseudo}^{rot}(\varphi,p)+\Psi_{pseudo}^{div}(\varphi,p) \qquad (5)$$

Because the divergence of the rotational horizontal velocity is zero ($\frac{1}{a\cos\varphi}\left(\frac{\partial u^{rot}}{\partial\lambda}+\frac{\partial v^{rot}\cos\varphi}{\partial\varphi}\right)=0$), the vertical velocity associated with rotational part of meridional $\Psi_{pseudo}$ is vertically uniform ($\frac{\partial\omega^{rot}}{\partial p}=0$). **With the upper boundary condition ($\omega=0$), it is reasonable to assume there is no vertical velocity associated with the rotational part of pseudo streamfunction ($\omega^{rot}=0$).**

In Sections 4 and 5, both the pseudo-streamfunction and the velocity vector $\vec{V}=(v,-\omega)$ will be presented in the figures to assure the robustness of regional pseudo-streamfunction. Note again, the pseudo-streamfunction $\Psi_{pseudo}$ itself carries physical significance as it represents the downwardly-integrated meridional mass transport. Consequently, abrupt seasonal changes in $\Psi_{pseudo}$ may serve as a useful indicator of abrupt seasonal changes in the meridional mass circulation (transport).

**3.3 Abrupt Seasonal Change Index (ASCI)**

We define an Abrupt Seasonal Change Index (ASCI) for the ASC of a physical quantity ψ (zonal wind, streamfunction, *etc*., but mainly pseudo-streamfunction in this paper) within a given 3-D region P:

Define $\Omega = \frac{\overline{\psi_+} - \overline{\psi_-}}{2}$, in which $\overline{\psi_+}$ stands for the average value of ψ within P if ψ>0, and $\overline{\psi_-}$ stands for the average value of ψ within P if ψ<0.

If more than 65% area of ψ is >0 within P, then  (ASCI$_1$, ASCI$_2$) = (Ω, 0);

Else if more than 65% area of ψ is <0 within P, then  (ASCI$_1$, ASCI$_2$) = (-Ω, 0);

Else  (ASCI$_1$, ASCI$_2$) = (0, Ω).

This method effectively categorizes atmospheric circulation within P into three regimes, provided that physical property ψ accurately represents the atmospheric circulation (Figure 1b). For example, if ψ is zonal wind (*u*), and (ASCI$_1$, ASCI$_2$) = (Ω, 0), then the westerly dominates this area; if (ASCI$_1$, ASCI$_2$) = (-Ω, 0), then the easterly dominates; if (ASCI$_1$, ASCI$_2$) = (0, Ω), then the easterly and westerly is comparable in the area. Also, the absolute value |Ω| determines the strength of each regime. In this paper, we use the vertical layer between 300hPa and 700hPa and 25°S-25°N as P, and we use meridional pseudo-streamfunction ($\Psi_{pseudo}$) and its zonal-sector average as the physical quantity ψ. Figure 1b is a schematic diagram for the definition of ASCI.

The two components of ASCI may be used to represent three main regimes (states)

of tropical meridional circulation:

1. Ascending branch in northern hemisphere and descending branch in southern hemisphere ($ASCI_1<0$, as shown in Figure 1b (I));

2. Ascending branch in southern hemisphere and descending branch in northern hemisphere ($ASCI_1>0$, as shown in Figure 1b (II));

3. Nearly equatorially symmetric state ($ASCI_2>0$, as shown in Figure 1b (III)). Usually each of the three regimes may last for 2-6 months, and we may define abrupt seasonal change by the following criteria: (1) the two regimes/states before and after the abrupt change last at least for 1 month (six pentads); (2) the transition period between the two regimes/states is no longer than 2 pentads.

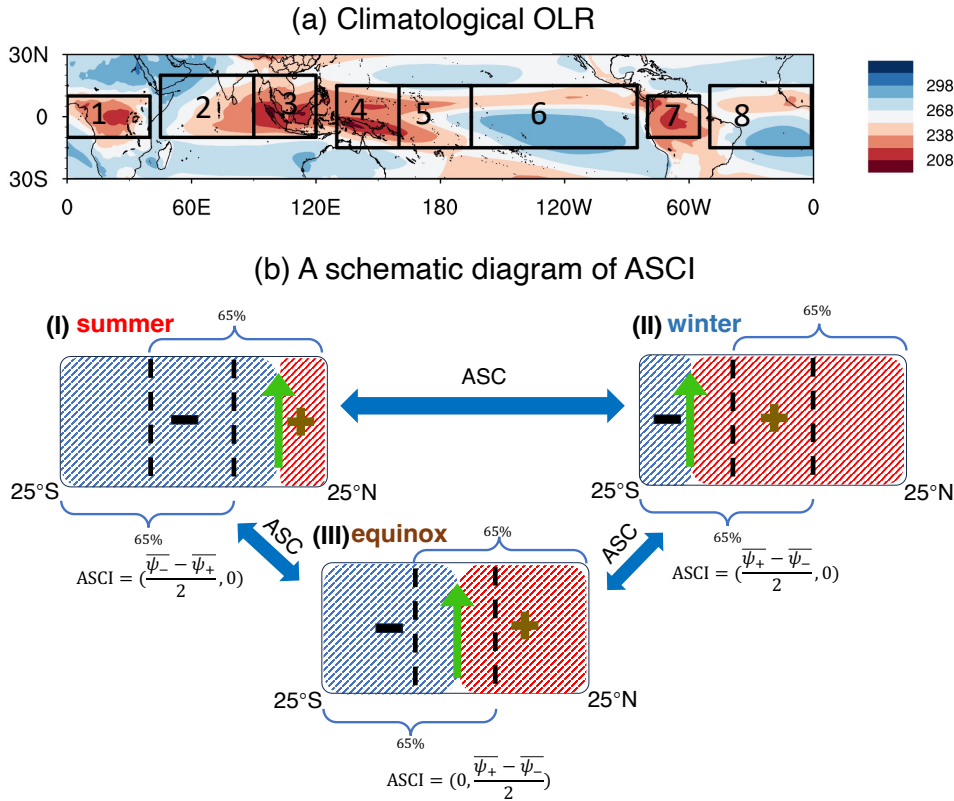

Figure 1 (a) Climatological annual mean of OLR obtained from CERES EBAF during 2000-2024. Unit: W/m$^2$. The numbers and the boxes indicate the regions that we use in this paper. Region 1-8 are the zonal sectors of Central Africa, Indian Ocean, South China Sea, the Maritime Continent, Central Pacific, Eastern Pacific, South America and Atlantic, respectively; (b) A Schematic diagram depicting the definition of Abrupt Seasonal Change Index (ASCI). The season labels are from the perspective of the northern hemisphere. (I) Boreal-summer-like meridional circulation, with $ASCI_1<0$; (II) Boreal-winter-like meridional circulation, with $ASCI_1>0$; (III) equinoctial circulation pattern, with $ASCI_2>0$. "ASC" in the figure stands for abrupt seasonal change between the three regimes. Green arrows indicate the rising branch of the tropical meridional circulation.

## 4 Regions with Abrupt Seasonal Change of Tropical Meridional Circulation

With the abovementioned definitions in Section 3, we are able to show the seasonal migration of ASCI in different regions. In this section, we first examine the ASC of meridional circulation over three regions associated with strong deep convection, namely, the tropical regions (25 °S-25 °N) over the zonal sectors of South China Sea (in the left panels of Figures 2 and 3), the Maritime continent (in the middle panels of Figures 2 and 3), and South America (in the right panels of Figures 2 and 3). Then we examine the ASCI in some other sub-regions in Figure 1a.

### 4.1 The zonal sector of South China Sea (Region 3)

Here we depict the climatological seasonality of ASCI from the 1st pentad (days 1-5 of a year) to the 73rd pentad (days 361-365 of a year).

In Figure 2a, we observe that the seasonal cycle over the zonal sector of South China Sea can be divided into two distinct regimes: one between pentads 26-59 (similar to the pattern of Figure 1b (I)) and another comprising pentads 60-73 and pentads 1-25 (similar to the pattern in Figure 1b (III)), referred to as the summer regime and the equinox regime, respectively (both from the perspective of the northern hemisphere, and hereafter). The transitions between the two regimes occur abruptly at pentad 59 (the late October) and at pentad 25 (the early May), with the ASCI showing abrupt transition between the summer regime (Figure 1b (I)) and the equinox regime (Figure 1b (III)). The timing of abrupt seasonal change from equinox to summer matches the time of monsoon outburst in South China Sea, which also indicates the beginning of East Asia Summer Monsoon (Wang et al., 2004).

The meridional flow during the summer regime over this region (Figure 2d) exhibits characteristics of a solstitial meridional circulation, with air ascending between 5°S and 30°N and descending between 30°S and 5°S. Conversely, during the equinox regime (Figure 2g), the meridional flow reflects an equatorially symmetric circulation, resembling the equinoctial Hadley Cell, with the ascending branch of the meridional circulation near the equator. Notably, the zonal wind structures differ significantly between the two regimes (see the gray contours in Figures 2d and 2g). While the ascending branches of the meridional circulation in both regimes exhibit westerlies near the surface, their meridional positions vary: for the summer regime, it ranges between 0° and 20°N; for the equinox regime, it ranges between 10°S and 5°N.

Obviously, the seasonal cycle of atmospheric circulation over this region fulfills the criteria for ASC. This can be further validated by examining consecutive pentads around the time of abrupt changes [Figures 3 a-e, representing the circulation patterns of this region during pentads 57-61 (mid-October to the end of a year), respectively]. The transition from the summer regime (Figure 3a, 57$^{th}$ pentad) to the equinox regime (Figure 3c, 59$^{th}$ pentad) occurs rapidly within just 2 pentads (10 days). Specifically, focusing on the meridional wind in 10°N-30°N between 100-300hPa, we find that northerly dominates before the 58th pentad, but the transition to southerly occurs abruptly after the 58th pentad. Accompanying the change in meridional wind, the meridional scope of the northern clockwise cell expands. In the 57th pentad (mid-October), the northern clockwise cell nearly does not exist yet, with the southern counter-clockwise cell dominating. However, by the 59th pentad (late-October), the maximum streamfunction for the northern cell has increased significantly, nearly matching that of the southern cell. By the 61st pentad (end-of-October), the maximum streamfunction of the northern cell surpasses that of the southern cell, approaching $1.5 \times 10^5$ Pa·m/s. The abrupt change from the winter regime into the summer regime bears similar processes, with the upper-level wind in 10°N-30°N changing sign between 26th-28th pentad (early-to-mid May, Figure not shown). The ascending branch of the Hadley cell also expands from 0-10°N to 0-30°N.

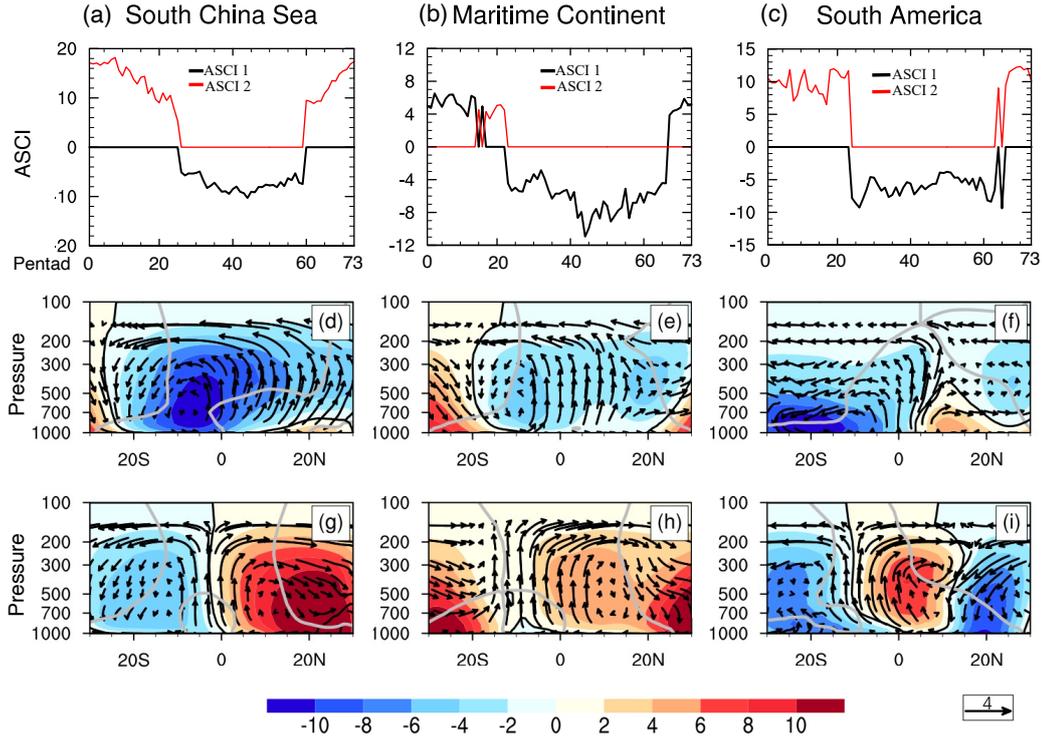

Figure 2 (a-c) Seasonal evolution of ASCI using ERA5 (unit: $10^4$ m/s*Pa, the black line is $ASCI_1$, while the red line is $ASCI_2$) over the zonal sectors of South China Sea (a), the Maritime Continent (b), and South America (c). (d-i) are the corresponding streamfunction (shading, unit: $10^4$ Pa*m/s) and velocity vector (v, -50*ω) (unit: m/s, Pa/s) during summer and winter: (d) summer (pentad 26-59) and (g) winter (pentad 60-73, plus 1-25) averaged over South China Sea; (e) summer (pentad 23-66) and (h) winter (pentad 67-73, plus pentad 1-15) averaged over the Maritime Continent; (f) austral winter (pentad 24-63) and (i) austral summer (pentad 64-73, plus pentad 1-23) averaged over South America. Gray contours represent the 0m/s zonal wind contour, while black contours represent a 0 Pa*m/s value of the streamfunction. The bottom abscissa of (a-c) shows an annual cycle from the 1st pentad to 73rd pentad.

## 4.2 The zonal sector of the Maritime Continent (Region 4)

The meridional circulation over the Maritime Continent represents a unique case regarding the seasonal migration of ASCI. Abrupt seasonal change is salient because ASCI jumps (Figure 2b, pentad 66-67, i.e., late-November) from the summer regime (Figure 2e, similar to Figure 1b (I)) into the winter regime (Figure 2h, similar to Figure 1b (II)) directly and abruptly, with a large jump in the area that the northern (southern) cell covers, i.e. from over 65% (less than 35%) to less than 35% (over 65%) abruptly in the $66^{th}$-$67^{th}$ pentad. However, the transition of ASCI from the winter regime to the summer regime is not a mirror to the transition from summer to winter. It first turns (Figure 2b, pentad 16-17, mid-to-late March) from the winter regime (Figure 1b (II)) into the equinox regime (Figure1b (III)) and then turns (Figure 2b, pentad 23-24, late-April) into the summer regime (Figure 1b (I)).

The result that the ASC over the Maritime Continent is asymmetric between the transition of summer to winter and the transition of winter to summer resembles the idealized GCM results of Geen et al. (2020, Figure 9), in which the seasonal cycle of ITCZ was shown with a prescribed geographical asymmetry between land and ocean. When the continent has southern boundaries at 0°N, the seasonal migration of precipitation maxima shows abrupt jump from 20°S to 0°N in March, while the reverse process involves the emergence of two maxima in October.

Figures 3f-j illustrate the consecutive pentads of pentad 65-69 (the late November to the early December), during which the ASC from the summer regime to the winter regime happens over the Maritime Continent region. In pentad 65-66, the summer regime has considerable strength (cool-tone shading), with the upper troposphere of 20°S-20°N dominated by northerlies. However, just after the transition period (pentad 67), the circulation is dominated by winter regime, while upper-level southerly prevails between 30°S-30°N nearly everywhere (except 5°S-10°S) from the $68^{th}$ pentad onwards (the vectors in Figures 3 i-j).

### 4.3 The zonal sector of South America (Region 7)

Figure 2c shows that the seasonal cycle of the meridional circulation over the South America can be divided into two regimes, in which the austral winter regime features one counter-clockwise circulation pattern (cool-tone shading, Figure 2f) with the ascending branch of the Hadley cell located between 5 °N and 10 °N, and the austral summer regime features a triple-cell structure (Figure 2i).  Focusing on the ASCI in Figure 2c, we find that the transition from the austral summer regime to the austral winter regime is abrupt in $23^{rd}$-$24^{th}$ pentads (late-April). However, oscillation occurs during the transition from "austral winter" to "austral summer" around the $63^{rd}$ pentad (early-November).

Before the ASC (Figures 3k-l, pentad 22-23, i.e., the mid-to-late April), the circulation retains in the winter regime as the abovementioned triple-cell structure (Figure 2i). However, in the $24^{th}$ pentad (the late April, Figure 3m) the clockwise circulation (warm-tone shading, 0°N-15°N) shrinks abruptly. Simultaneously, there is a decrease in the strength of the counterclockwise circulation (cool-tone shading, 0°S-30°S) from over $15*10^4$ Pa*m/s to about $9*10^4$ Pa*m/s. Meanwhile, the northern cell (20°N-30°N) strengthens in the $24^{th}$ pentad. After the $24^{th}$ pentad, although there are changes in the strength of each cell, their meridional span and strength never revert to their counterparts in the winter regime.

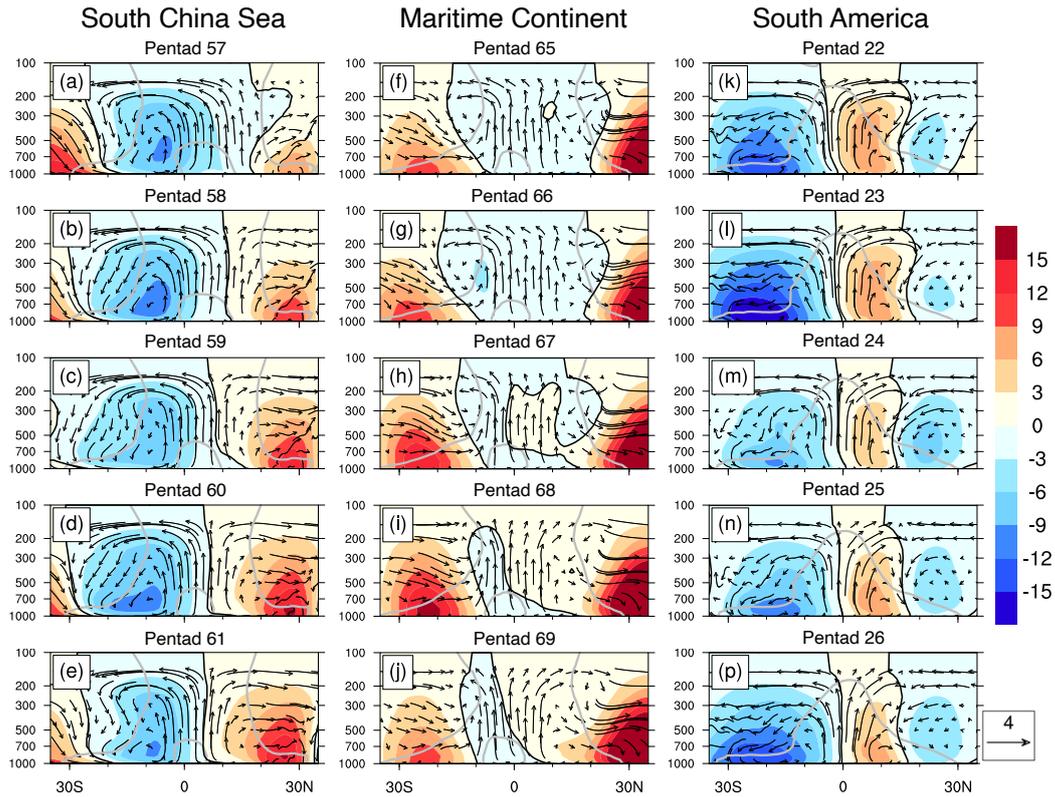

Figure 3 Consecutive pentads of meridional flow around the time abrupt seasonal change happens according to ASCI using ERA5. (a-e) is pentad 57-61 of South China Sea; (f-j) is pentad 65-69 of the Maritime Continent; (k-p) is pentad 22-26 of South America. The definition of the shading, the vector and the contour are the same as in Figure 2(d).

## 4.4 Other Regions

At first glance, there are no clear signals (Figure S1) of abrupt seasonal change in regions such as the Africa (region 1), the eastern Pacific (region 6), and the Atlantic (region 8). Indeed this may be true for some regions such as a vast zonal range over the middle to eastern Pacific (Adam et al., 2016). The signal of ASC, however, may be blurred by zonally averaging over a longitudinal range that is so large that contains different subsystems – similar to the case of Hadley cell zonally averaged over the whole zonal circle (Dima and Wallace, 2003). Therefore, a more careful scrutiny is needed for the ASC in these regions.

The meridional circulation within a limited longitudinal span cannot be independent from the horizontal flow pattern and meridional overturning circulation within a much larger longitudinal range. Take Africa (region 1 in table S1) as an example, African Easterly Jet (AEJ) dominates the lower troposphere of Western Africa and Central Africa (Nicholson and Grist, 2003), linking the Azores High (North Atlantic High) and South Atlantic High, significantly influencing local climate. On the other hand, the Somali Jet dominating the eastern part of Africa is closely related to the Indian Monsoon system (Krishnamurti et al., 1976).

By taking these into consideration we subdivide the regions into smaller ones, and then we find that the ASCIs in some sub-regions show clearly the signal of abrupt seasonal change (Figure 4 a-d). Generally, abrupt seasonal change occurs either in the transition process from the summer regime to the equinox regime (Western Africa 0°E-10°E, Figure 4a and Atlantic 0°W-15°W, Figure 4d), or in the transition process from the equinox regime to the summer regime (Central Africa 10°E-20°E, Figure 4b), or both (Eastern Pacific 85°W-105°W, Figure 4c). Take the subregion of Eastern Pacific (85°W-105°W) as an example. From Figure 4g, we find that the summer regime features ascending motion in the northern hemisphere (5°N-20°N), and descending motion in the southern hemisphere (0°S-25°S). The equinox regime (Figure 4k) features a quasi-symmetric circulation pattern characterizing the asending branch near 5°S. We can testify the ASC indicated by ASCI (Figure 4c) by showing the evolution of circulation during the pentads around the pentad that abrupt seasonal change happens (pentad 58-62, the late October to the early November, Figure S2 k-p). Before pentad 58-59, the southern counterclockwise cell (cool-tone shading) dominates the tropical region (Figure S2 k-l), covering an area over 65% of the P area. Then the northern cell suddenly expands to nearly 0°N during the $60^{th}$ pentad (Figure S2m) and the southern cell suddenly shrinks, indicating the onset of the equinox regime. The transition from the equinox to the summer regime also shows abrupt change clearly (Figure not shown). Details of abrupt seasonal change in other subregions have been shown in the Supporting Information.

In short, nearly all regions show abrupt seasonal change, either globally or locally. However, the areas with strong deep convection show ASC more profoundly, this might be not only due to the more active nature of the atmosphere convection in these regions and the land-sea configuration (Geen et al., 2020), but also due to the feedback between large-scale extratropical eddies and the tropical circulation as suggested by Bordoni and Schneider (2008).

In addition, we use JRA-55 datasets (Figure S3-S4) as a validation of the results above. Signals of ASC in JRA-55 are nearly the same as in ERA5, although the intensity of ASCI exhibited by the two datasets occasionally differs.

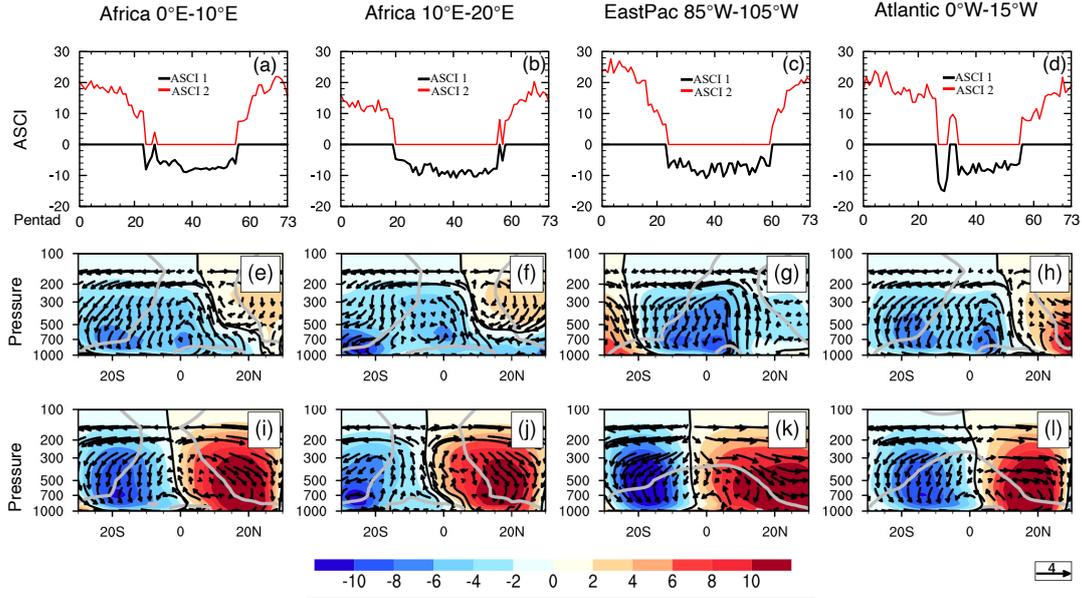

Figure 4 The same as Figure 2, but (a-d) depict the seasonal evolution of ASCI using ERA5 of (a) Western Africa, 0°E-10°E, (b) Central Africa, 10°E-20°E, (c)Eastern Pacific, 85°W-105°W, (d)Atlantic, 0°W-15°W. (e-l) are the corresponding streamfunction (shading, unit: $10^4$ Pa*m/s) and velocity vector (v, -50*ω) (unit: m/s, Pa/s) during summer and winter: (e) boreal summer and (i) boreal winter averaged over 0°E-10°E of Central Africa; (f) boreal summer and (j) boreal winter averaged over 10°E-20°E of Central Africa; (g) boreal summer and (k) boreal winter averaged over 85°W-105°W of Eastern Pacific; (h) boreal summer and (l) boreal winter averaged over 0°W-15°W of Atlantic.

## 5 Rotational Flow Dominates the regional Abrupt Seasonal Changes

We emphasize that the ASC of zonally asymmetric Hadley cell may be dynamically different from the ASC of zonally symmetric Hadley cell. Recently Hoskins and Yang (2023) proposed that the dynamics of zonally asymmetric Hadley cell, particularly that of the upper-branch angular momentum balance and vorticity budget, may involve the tropical eddies associated with tropical deep convection. Similarly the quasi-stationary and transient tropical eddies, including tropical Rossby waves and mixed gravity-Rossby waves, can be playing essential roles in the ASC of regional and zonally asymmetric Hadley cell. Here we quantitatively compare the contribution of rotational flow (usually associated with Rossby waves or mixed gravity-Rossby waves) and divergent flow (usually associated with gravity modes and thermally direct response to convection) to the regional, zonally asymmetric ASC.

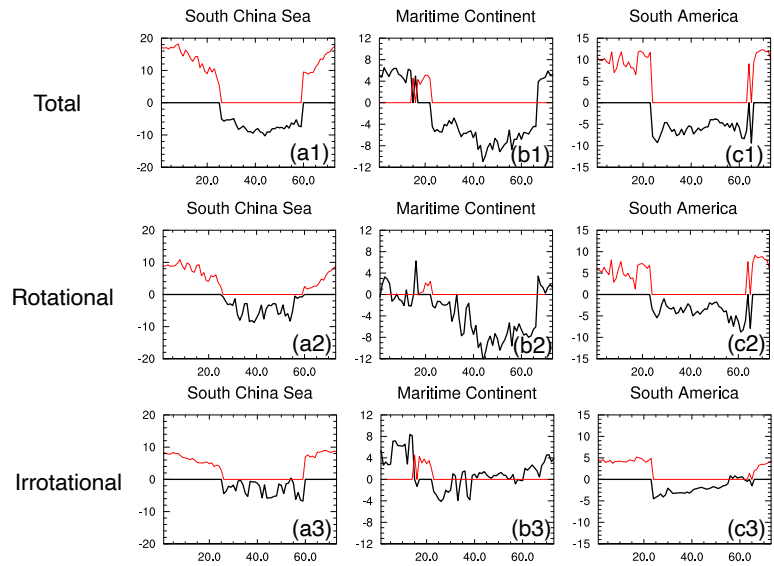

Figure 5. The decomposition of ASCI (upper panel) into the sum of rotational parts (middle panel) and divergent parts (bottom panel) for the zonal sectors of South China Sea, the Maritime Continent, and South America.

In Figure 5, we decompose the total ASCI over the zonal sectors of South China Sea (a1), the Maritime Continent (b1), and South America (c1) into the sum of rotational parts (middle panel, a2-b2-c2) and divergent parts (bottom panel, a3-b3-c3). While the divergent components may partly contribute to the ASC, the contributions from the rotational flow dominate the zonally asymmetric ASC over the three zonal sectors containing deep convection. The rotational contribution is even more dominant over the four sub-regions (Figure 6) corresponding to that in Figure 4.

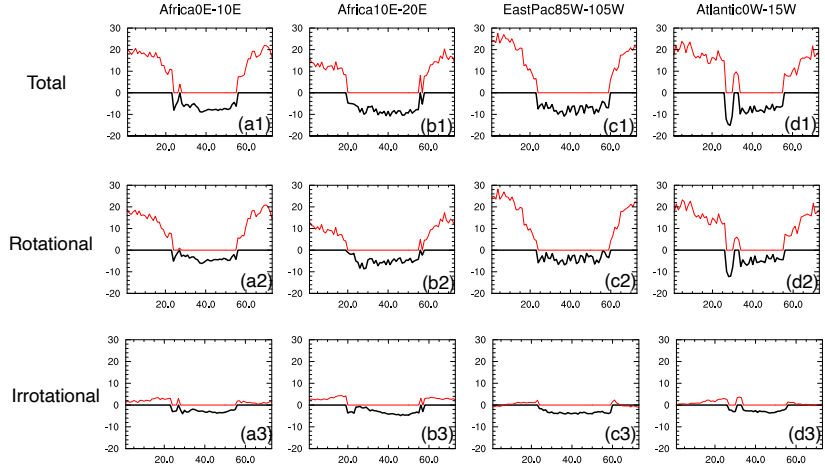

Figure 6. The same as Figure 5, but for the zonal sectors of (a) western Africa (0°E-10°E, a1-a3); (b) Central Africa (10°E-20°E, b1-b3); (c) Eastern Pacific (85°W-105°W, c1-c3); (d) Atlantic (0°W-15°W, d1-d3), corresponding to that in Figure 4.

We may further examine the evolution of pseudo-streamfunction and its rotational and divergent parts during the abrupt transition periods. Figure 7 presents an example for the transition from summer regime to equinox regime in the zonal sector of South China Sea. The left panel (Figures 7 a-e) is the same as in Figures 3 a-e, and the middle and right panels are the rotational and divergent components of pseudo-streamfunction ($\Psi_{pseudo}$). We can see that the divergent part of $\Psi_{pseudo}$ (the right panel) gradually moves southward during this period with very little change in its strength, but there are salient, and sudden change in both the strength and location of the rotational component of $\Psi_{pseudo}$, which clearly contribute to the ASC.

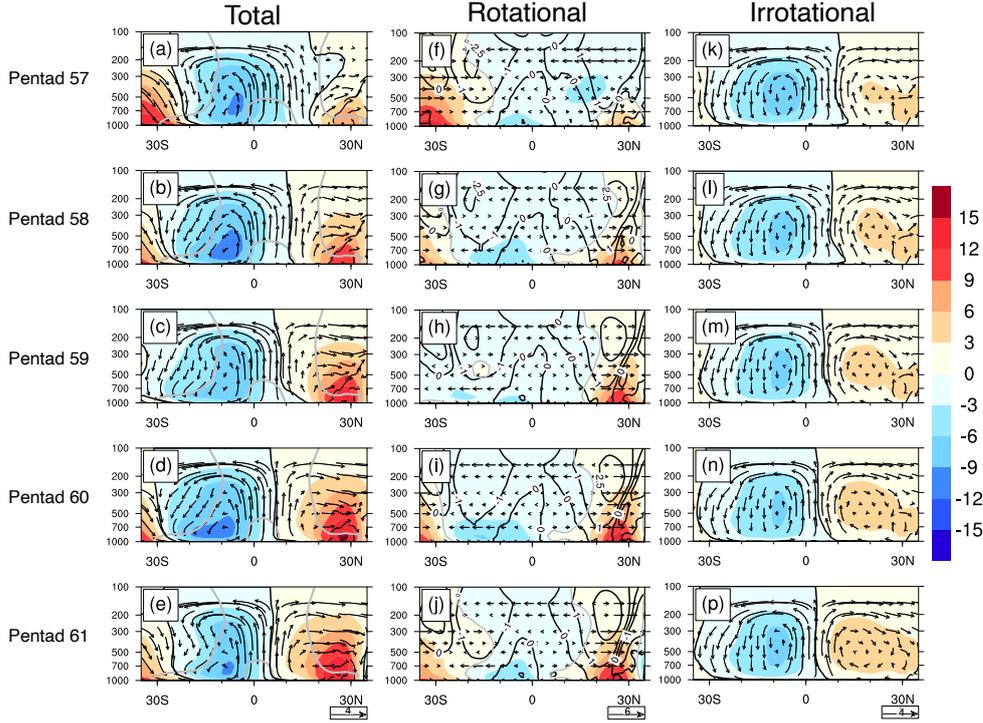

Figure 7. The evolution of pseudo-streamfunction $\Psi_{pseudo}$ (the left panel) its rotational part (middle panel) and divergent part (right panel) during the abrupt seasonal change from summer regime to equinox regime (around from pentad 57 to 61) over the zonal sector of South China Sea.

Indeed, similar cases are found for the transition from equinox regime to summer regime over the zonal sector of South China Sea (Figure 8), and also all of the ASCs over other regions mentioned above. Therefore, the dominant role of rotational flow in the regional, zonally asymmetric ASC of tropical meridional circulation is a robust feature. This is quite different from the zonally symmetric case in which the meridional circulation (streamfunction) is divergent. Bordoni and Schneider (2008) indicated, from a perspective of zonally homogeneous aquaplanet, that the feedback between large-scale extratropical eddy-induced momentum flux and tropical circulation, together with the small surface inertia, determines the ASC of monsoon circulation. While their conclusions are physically essential, the location-dependence of ASC of zonally asymmetric Hadley circulations, and particularly the dominant role of rotational flow in ASC suggest the interplay between tropical Rossby-type eddies and extratropical eddies may also be essential to the ASC of tropical circulation.

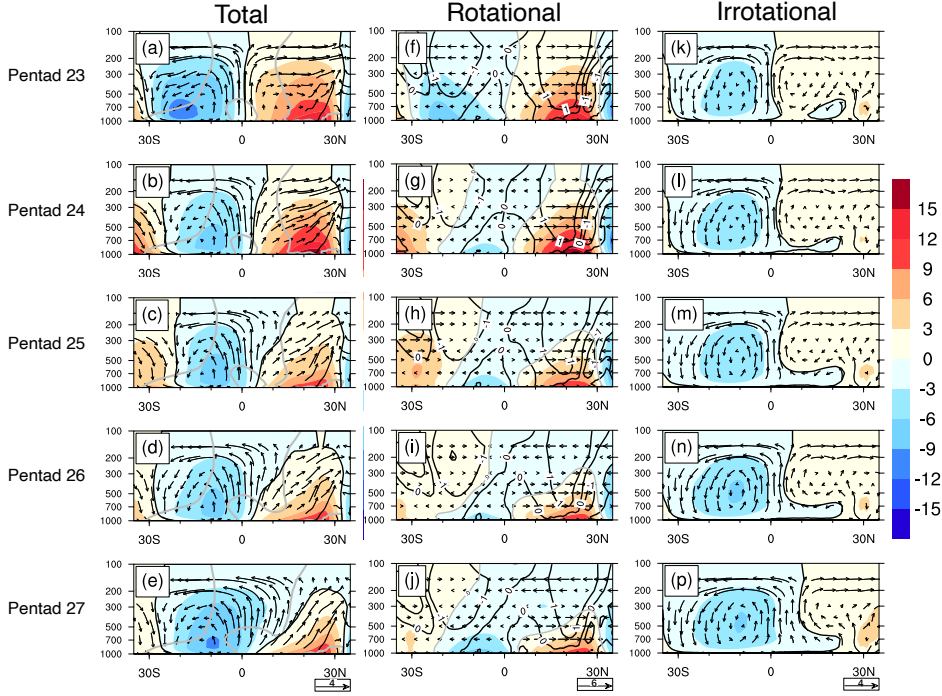

Figure 8. The same as Figure 7, but for the abrupt transition from equinox regime to summer regime over the zonal sector of South China Sea.

## 6 Conclusions and Discussion

We in this paper develop a systematic method to diagnose zonally asymmetric abrupt seasonal change (ASC) in tropical meridional circulation. We first define a pseudo-streamfunction ($\Psi_{pseudo}$) at any zonal sector, which is meant by the downwardly integrated meridional mass transport, and converges to traditional meridional overturning streamfunction ($\Psi$) after being averaged over a zonal circle around any latitude. We then define a new vector-type, dual-component index (Abrupt Seasonal

Change Index, ASCI). By applying Helmholtz decomposition to horizontal velocity fields, so as to decompose $\Psi_{pseudo}$ into the sum of its rotational and divergent parts, we quantitatively assess the contributions of horizontally rotational and divergent flows to the ASC of tropical circulation. In division of subregions in the tropics, we use outgoing longwave radiation (OLR) intensity as the criterion.

We find that the ASCIs over the zonal sectors, containing mainly the South China Sea, the Maritime continent and South America, all associated with strong deep convection, exhibit the most pronounced abrupt seasonal change, while ASCI in subregions of Central Africa, Eastern Pacific and Atlantic also show abrupt seasonal changes.

We also find that all of subregions exhibiting salient abrupt seasonal change are located in the regions having landmass with low heat inertia. In contrast, zonal sectors containing only with vast open oceans tend to show no sign of ASC, like the central Pacific and the western part of Eastern Pacific. This confirms that the low thermal inertia of land is essential to the ASC of tropical circulation (Bordoni and Schneider, 2008).

Most importantly, we find that the rotational flow, rather than the thermal-direct divergent flow, dominates the zonally asymmetric abrupt seasonal change in the tropics, although the latter also contributes to the ASC over zonal sectors associated with deep convection.

We emphasize that the ASC of zonally asymmetric Hadley cell may be dynamically different from the ASC of zonally symmetric Hadley cell. Recently Hoskins and Yang (2023) proposed that the dynamics of zonally asymmetric Hadley cell, particularly that of the upper-branch angular momentum balance and vorticity budget, may involve the tropical eddies associated with tropical deep convection. The dominant role of rotational flow in the ASC suggests that except large-scale extratropical eddies, the quasi-stationary and transient tropical eddies, including tropical Rossby waves and mixed Rossby-gravity waves, can be essential in the ASC of regional and zonally asymmetric Hadley cell.

**Acknowledgements**

The authors are supported by the National Natural Science Foundation of China (grant nos. 42175070 and 42288101).

**Author Contributions:** ASCI developed mainly by W. Yao with the suggestions from J. Lu; Helmholtz decomposition and rotational contribution originated from J. Lu's idea; Diagnostics performed by W. Yao; Result analysis: W. Yao and J. Lu; Conceptualization: J. Lu; Manuscript Writing: W. Yao and J. Lu; review and editing: W. Yao, J. Lu, and Y. Liu; Computing Resource and supervising: Y. Liu and J. Lu.

**Data Availability Statement**

The ERA5 datasets (Hersbach et al., 2020) from 1979 to present are available at: https://cds.climate.copernicus.eu/cdsapp#!/dataset/reanalysis-era5-pressure-levels?tab=form

The JRA-55 datasets (Kobayashi et al., 2015) from 1979 to present are available at: https://thredds.rda.ucar.edu/thredds/catalog/aggregations/g/ds628.0/3/catalog.html?dataset=aggregations/g/ds628.0/3/TP

The OLR dataset from Clouds and the Earth's Radiant Energy System Energy Balanced And Filled (CERES EBAF) (EBAF, Loeb et al., 2001) from 2000 to 2024 is available at: https://cds.climate.copernicus.eu/cdsapp#!/dataset/satellite-earth-radiation-budget?tab=form

# Supporting Information

|   | Geographical region | Short Name | Longitudinal range | Region type |
|---|---|---|---|---|
| 1 | Central Africa | CAf | 0°-40°E | Strong convection |
| 2 | Indian Ocean | IO | 45°E-90°E | Transition |
| 3 | South China Sea | SCS | 90°E-115°E | Strong convection |
| 4 | Maritime Continent | MC | 130°E-160°E | Strong convection |
| 5 | Central Pacific | CP | 160°E-165°W | Transition |
| 6 | Eastern Pacific | EP | 85°W-165°W | Weak convection |
| 7 | South America | SAm | 55°W-80°W | Strong convection |
| 8 | Atlantic | Atl | 0°-50°W | Weak convection |

Table S1 Geographical areas analyzed.

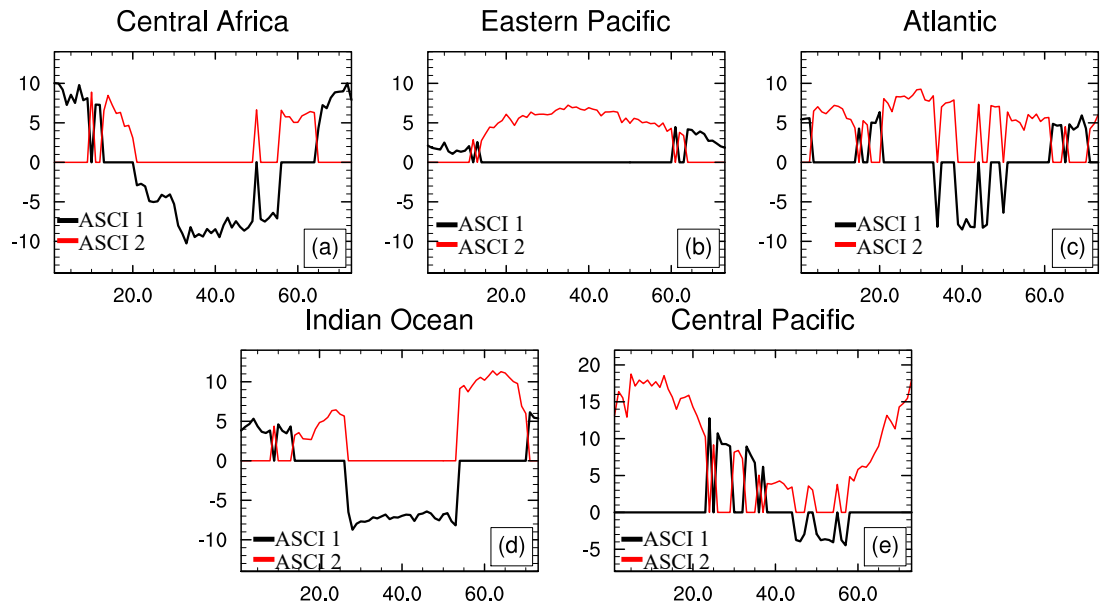

Figure S1 Seasonal evolution of ASCI using ERA5 (defined in Data and Methods, unit: $10^4$ m/s*Pa, the black line is $ASCI_1$, while the red line is $ASCI_2$) of (a)Central Africa, (b)Eastern Pacific, (c)Atlantic, (d)Indian Ocean, (e)Central Pacific.

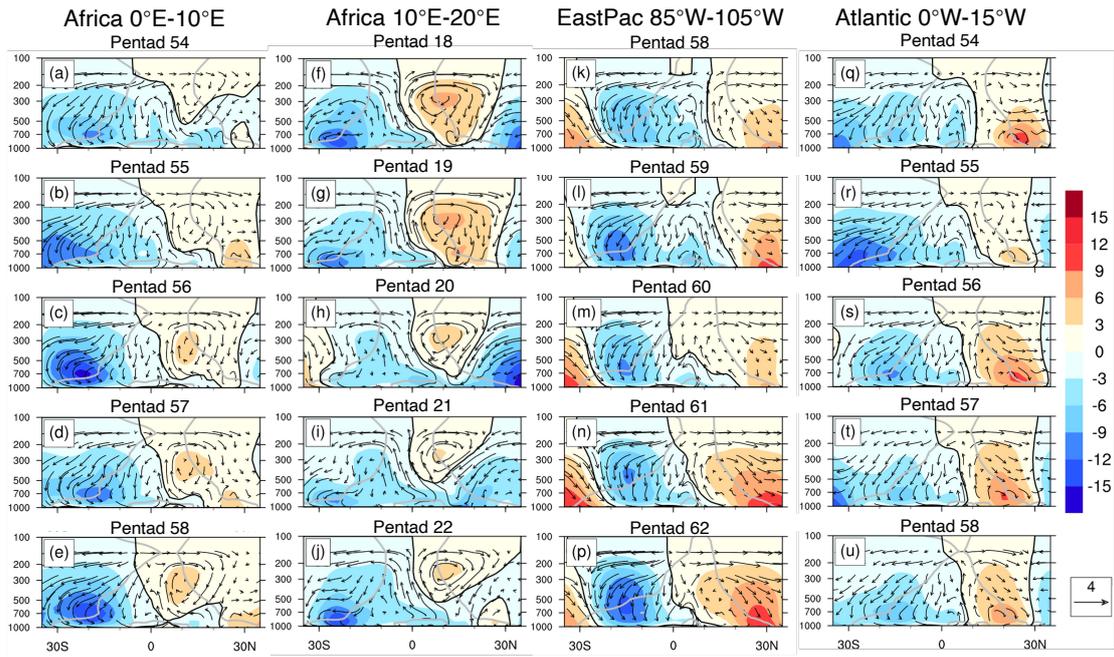

Figure S2 Consecutive pentads of meridional flow in the vicinity of abrupt seasonal change of different areas according to ASCI using ERA5. (a-e) is pentad 54-58 of Africa, 0°E-10°E; (f-j) is pentad 18-22 of Africa, 10°E-20°E; (k-p) is pentad 58-62 of Eastern Pacific, 85°W-105°W; (q-u) is pentad 54-58 of Atlantic, 0°W-15°W.

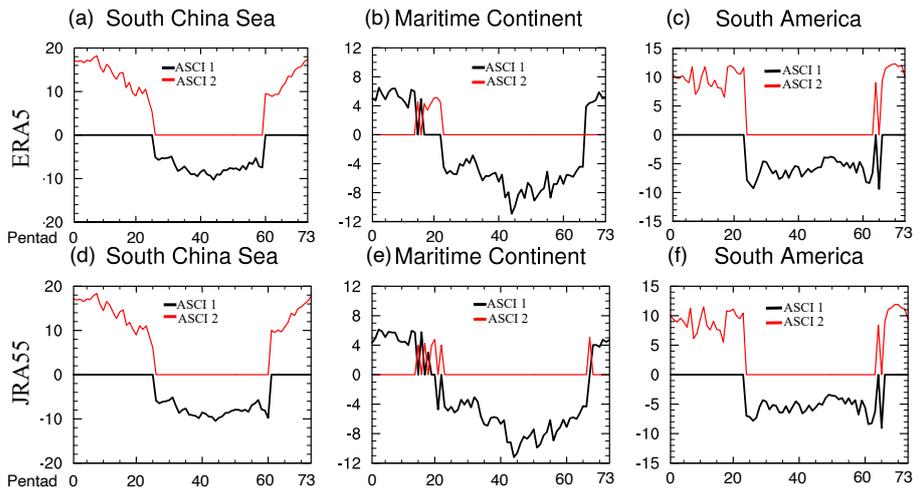

Figure S3 Seasonal evolution of ASCI using ERA5 (defined in Data and Methods, unit: $10^4$ m/s*Pa, the black line is $ASCI_1$, while the red line is $ASCI_2$) of (a)South China Sea, (b)Maritime Continent, (c) South America. (d-f) are as in (a-c) but using JRA-55.

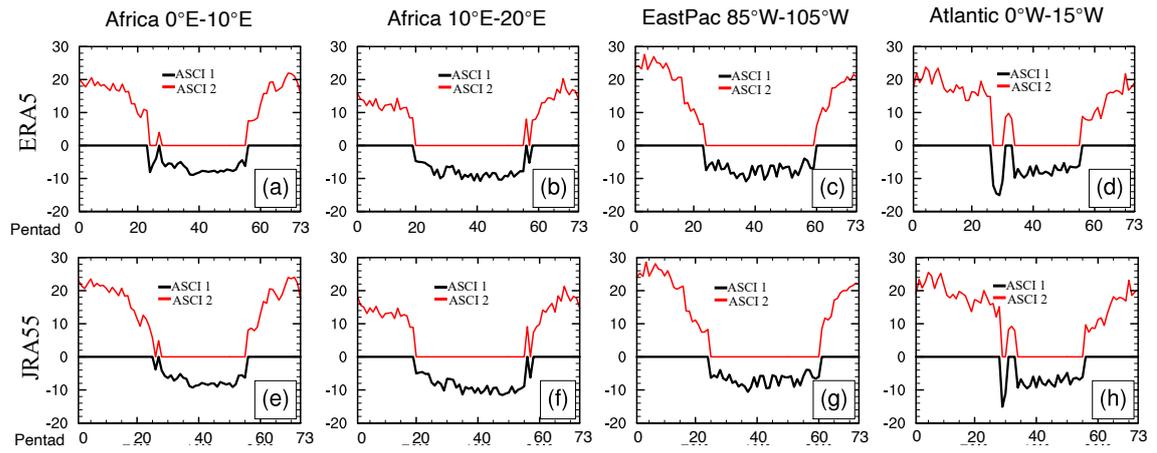

Figure S4 Seasonal evolution of ASCI using ERA5 (defined in Data and Methods, unit: $10^4$ m/s*Pa, the black line is $ASCI_1$, while the red line is $ASCI_2$) of (a) 0°E-10°E, (b) 10°E-20°E, (c) 85°W-105°W, (d) 0°W-15°W. (e-h) are as in (a-d) but using JRA-55.